\documentclass{article}

\PassOptionsToPackage{square,sort&compress,numbers}{natbib}
\usepackage[final]{neurips_2024}
\usepackage{fontspec}
\usepackage{xeCJK}
\setCJKsansfont{FandolHei-Regular.otf}[BoldFont=FandolHei-Bold.otf]
\setCJKmonofont{FandolFang-Regular.otf}[AutoFakeBold, AutoFakeSlant]

\usepackage{float}
\usepackage{wrapfig}
\usepackage{hyperref}
\hypersetup{
    colorlinks=true,
    linkcolor=blue,
    filecolor=magenta,
    urlcolor=blue,
    citecolor=blue,
    pdfinfo={
        Title={Certifying Foundation Models for Professional Education},
        Subject={foundation models, professional education, teacher certification, benchmarks},
        Keywords={EDU 1.0, Educational Due Diligence, foundation models, teacher certification, benchmark, professional education},
    }
}

\usepackage{url}
\usepackage{booktabs}
\usepackage{amsfonts}
\usepackage{amsmath}
\usepackage{nicefrac}
\usepackage{microtype}
\usepackage{xcolor}
\usepackage{graphicx}
\usepackage{soul}
\usepackage{cleveref}
\usepackage{enumitem}
\usepackage{multirow}
\usepackage{colortbl}
\usepackage{makecell}
\usepackage{amssymb}
\usepackage[most]{tcolorbox}
\usepackage{tabularx}

\definecolor{cardcn}{HTML}{4D97CD}
\definecolor{cardus}{HTML}{02AFAC}
\definecolor{cardfr}{HTML}{B379B4}
\definecolor{cardin}{HTML}{8B96AD}
\definecolor{carderr}{HTML}{DB6968}
\definecolor{cardbg}{HTML}{F8F8F8}

\newtcolorbox{qcard}[2][]{
  enhanced, boxrule=0.8pt, arc=2pt,
  left=6pt, right=6pt, top=4pt, bottom=4pt,
  colback=cardbg, colframe=#2,
  fonttitle=\bfseries\small, coltitle=white,
  attach boxed title to top left={yshift=-2mm, xshift=4mm},
  boxed title style={colback=#2, arc=1pt, boxrule=0pt},
  #1
}

\newtcolorbox{errcard}[1][]{
  enhanced, boxrule=0.8pt, arc=2pt,
  left=6pt, right=6pt, top=4pt, bottom=4pt,
  colback=#1!5!white, colframe=#1!80!black,
  fonttitle=\bfseries\small, coltitle=white,
  #1
}

\let\svthefootnote\thefootnote
\newcommand\freefootnote[1]{  \let\thefootnote\relax  \footnotetext{#1}  \let\thefootnote\svthefootnote}

\title{Measuring the Professional Educational Competence of Foundation Models}
\author{  Keqian Li\textsuperscript{4} \quad
  Mingzi Zhang\textsuperscript{2} \quad
  Xiaolong Wang\textsuperscript{4} \quad
  Aimin Zhou\textsuperscript{1,3}\\[3pt]
  \normalfont\footnotesize
  \textsuperscript{1}Shanghai Innovation Institute \quad
  \textsuperscript{3}East China Normal University\\
  \normalfont\footnotesize
  \textsuperscript{2}Faculty of Education at East China Normal University (ECNU)\\
  \normalfont\footnotesize
  \textsuperscript{4}Shanghai Institute of AI for Education, East China Normal University (ECNU)\\[3pt]
  \normalfont\footnotesize\textbf{Website:} \url{https://ecnu-innospark.github.io/EDU}\\
  \normalfont\footnotesize\textbf{Code:} \url{https://github.com/ECNU-innoSpark/EDU}\\
  \normalfont\footnotesize\textbf{Data:} \url{https://huggingface.co/datasets/keqianli/EDU}\\
}
\begin{document}

\newcommand{\fullname}{\textsc{EDU 1.0}}
\newcommand{\name}{\textsc{EDU}}
\newcommand{\questioncount}{$7{,}496$}

\newcommand{\cilevel}{95}
\newcommand{\ci}[3]{#1\%\ (\cilevel\%~CI,\ #2--#3)}
\newcommand{\cis}[3]{#1\%\ (#2--#3)}

\maketitle

\freefootnote{\hspace{-1em}Complete author information will be added in the camera-ready version.}

\renewcommand{\thefootnote}{\fnsymbol{footnote}}

\renewcommand{\thefootnote}{\fnsymbol{footnote}}

\begin{abstract}

Education has become one of the most consequential arenas for AI deployment: foundation models now tutor, assess and instruct at population scale, making it essential to quantify their progress against rigorous, externally defined standards.
Policy calls for validation before classroom use, yet existing benchmarks do not assess models against authentic teacher-entry standards: general evaluations emphasize difficult academic problems, whereas education-specific benchmarks often rely on isolated, synthetic tasks.
We introduce EDU 1.0 (Educational Due Diligence for Foundation Models), which uses teacher-entry assessments as externally defined proxies for educational competence rather than substitutes for human teacher qualification.
EDU 1.0 comprises 10,012 questions from teacher certification and recruitment examinations including the U.S. Praxis series, China's National Teacher Qualification Examination, and India's Kendriya Vidyalaya Sangathan examinations, covering foundational literacy and knowledge, pedagogical principles, and subject-specific pedagogical expertise across language arts, mathematics, science, social science, and education practice and student development.
Across 36 foundation model variants, capable educational performance is no longer
confined to proprietary systems: the strongest proprietary model reaches a
response-balanced score of 96.5\%, the leading open-weight model trails by 2.1
percentage points, and the leading system deployable on a single accelerator
reaches 92.2\%.
These aggregate scores conceal where the systems still differ: every one of them
scores higher on general pedagogical principles than on the subject assessments
that require applying that pedagogy within a discipline, and across those
assessments each system's own scores span 4.1 to 8.8 points while its shortfall
relative to general pedagogy widens from 2.6 to 4.1 points as capability
declines.
What remains is therefore pedagogical content knowledge rather than
pedagogy or scale. These results therefore locate the outstanding requirement in pedagogical content
knowledge---the capacity to make particular subject matter teachable to particular
learners---rather than in general pedagogical knowledge or in model scale.

\end{abstract}

\begin{figure}[H]
    \centering
        \includegraphics[width=\textwidth]{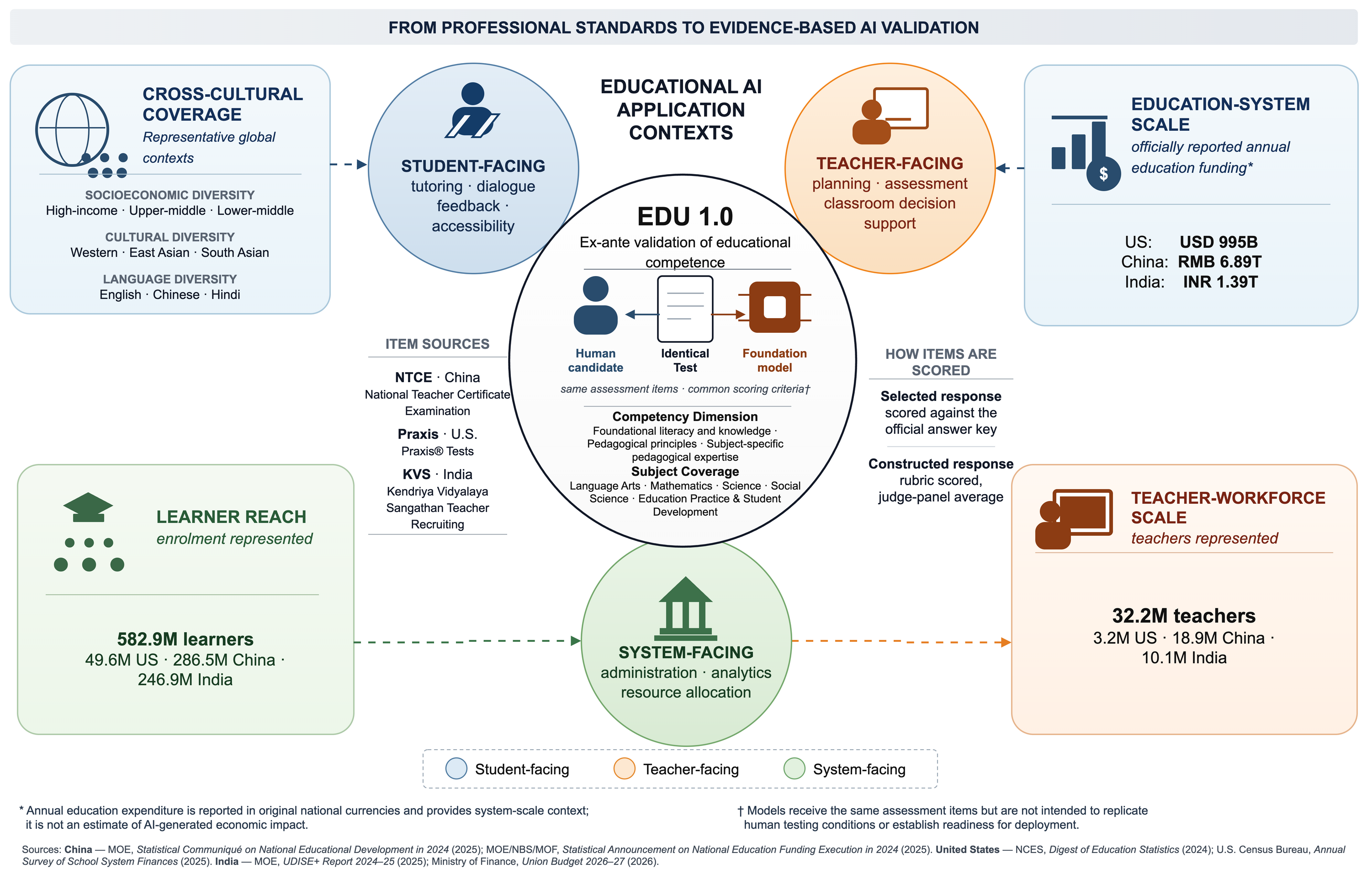}
    \caption{
        \textbf{EDU 1.0 as a profession-aligned evaluation framework.}
        \textbf{a}, The three teacher-entry systems represented in EDU 1.0---the U.S.\ Praxis, China's NTCE, and India's KVS recruitment examinations---with the students and teachers they serve and their annual education funding~\cite{moe2025communique, nces2024digest, udise2025report, moe2025funding, census2025schoolfinances, indiabudget2026}.
                        \textbf{b}, EDU 1.0 evaluates foundation models against the shared competency profile represented by these systems---foundational literacy and knowledge, pedagogical principles, and subject-specific pedagogical expertise---using 10,012 authentic items in written, interview, and demonstration formats. Models receive the same assessment items but are not intended to replicate human testing conditions or establish readiness for deployment.
                        \textbf{c}, The competencies evaluated map onto application contexts including AI tutors, teacher copilots, and standards-aligned assessment. These contexts illustrate the relevance of profession-aligned evaluation rather than downstream impact or deployment readiness.
    }
    \label{fig:edu_infrastructure}
\end{figure}

\section{Introduction}\label{sec:introduction}

There is now a global recognition, codified in UNESCO guidance and the Beijing Consensus, that artificial intelligence may reshape the very foundations of teaching and learning~\cite{miao2021policy,unesco2019beijing}.
From mechanical teaching machines to intelligent tutoring systems, educational technology has long sought to augment and extend aspects of human instruction~\cite{skinner1958teaching,miao2021policy}.
Generative foundation models mark a qualitative shift in educational AI by combining broad knowledge, multimodal interaction, and open-ended generation ``across all key symbolic representations of human thinking'', extending educational AI beyond narrowly defined functions~\cite{unesco2023genai}.
The McKinsey report decomposes teachers' work into instructional activities and surrounding labour---including preparation, evaluation and feedback, administration, record maintenance, and professional development---and estimates that existing technologies could potentially automate portions of this workload, corresponding to 20--40\% of total working time in some scenarios~\cite{bryant2020teachers}.
Beyond the teacher's desk, student-facing tutoring systems have been pursued since the 1970s, and system-facing tools now manage admissions, scheduling, and records~\cite{miao2021policy}, with potential impact amplified by sheer scale:
the United States, China, and India alone enrol over 580 million students and employ over 32 million teachers, with annual education spending at the trillion scale
(\Cref{fig:edu_infrastructure})~\cite{moe2025communique,nces2024digest,udise2025report,moe2025funding,census2025schoolfinances,indiabudget2026}.

More is at stake than efficiency: education anchors the 2030 Agenda, and UNESCO member states have pledged to harness AI for SDG 4---quality education for all~\cite{unesco2019beijing}.
Yet UNESCO warns that convincing but erroneous GenAI material ``poses a high risk for young learners who do not have solid prior knowledge of the topic in question''~\cite{unesco2023genai}.
In response, UNESCO calls for GenAI applications in education to undergo validation before institutional adoption, including assessment of safety, educational effectiveness, age appropriateness, and alignment with pedagogical principles~\cite{unesco2023genai}.
Such validation must consider not only what models can do, but also whether their capabilities are practically accessible to education systems operating under different resource constraints; international AI governance discussions increasingly emphasize equitable access to AI capabilities on precisely these grounds~\cite{unitednations2024gdc}.
Models whose deployment depends on proprietary access or industrial-scale infrastructure may remain inaccessible where computational resources are limited.
Existing evaluations have substantially advanced the measurement of model knowledge and specific educational interactions, but leave open whether models meet profession-aligned educational standards.

General evaluations such as MMMU~\cite{yue2024mmmu}, MMLU-Pro~\cite{wang2024mmlupro}, and HLE~\cite{phan2025hle} have advanced measurement on increasingly difficult academic problems and expert-level tasks. However, they primarily assess what models know or can solve, rather than whether they can support human learning in authentic educational contexts~\cite{unesco2023genai}; their task distributions can also differ substantially from the learners and curricula encountered in deployment~\cite{oecd2023equity}. This leaves open whether foundation models meet profession-aligned educational standards---evidence increasingly required for validation before educational deployment.
Education-specific benchmarks have substantially advanced evaluation of particular educational interactions~\cite{xu2025edubench,srinivasa2025tutorbench,yue2026edustudybench}.
EduBench, for instance, scores model responses to 198 LLM-synthesized query--response instances spanning nine educational scenarios ~\cite{xu2025edubench}
; its peers focus on mathematics tutoring dialogues~\cite{macina2025mathtutor}, hint-and-feedback tasks~\cite{srinivasa2025tutorbench}, or professional ethics probed through 88 constructed dilemmas~\cite{jiang2025moral}.
Each ranks models usefully within its intended scope, but none evaluates the breadth of competencies represented in professional teacher-entry standards.

\begin{figure}[H]
    \centering
    \includegraphics[width=\textwidth]{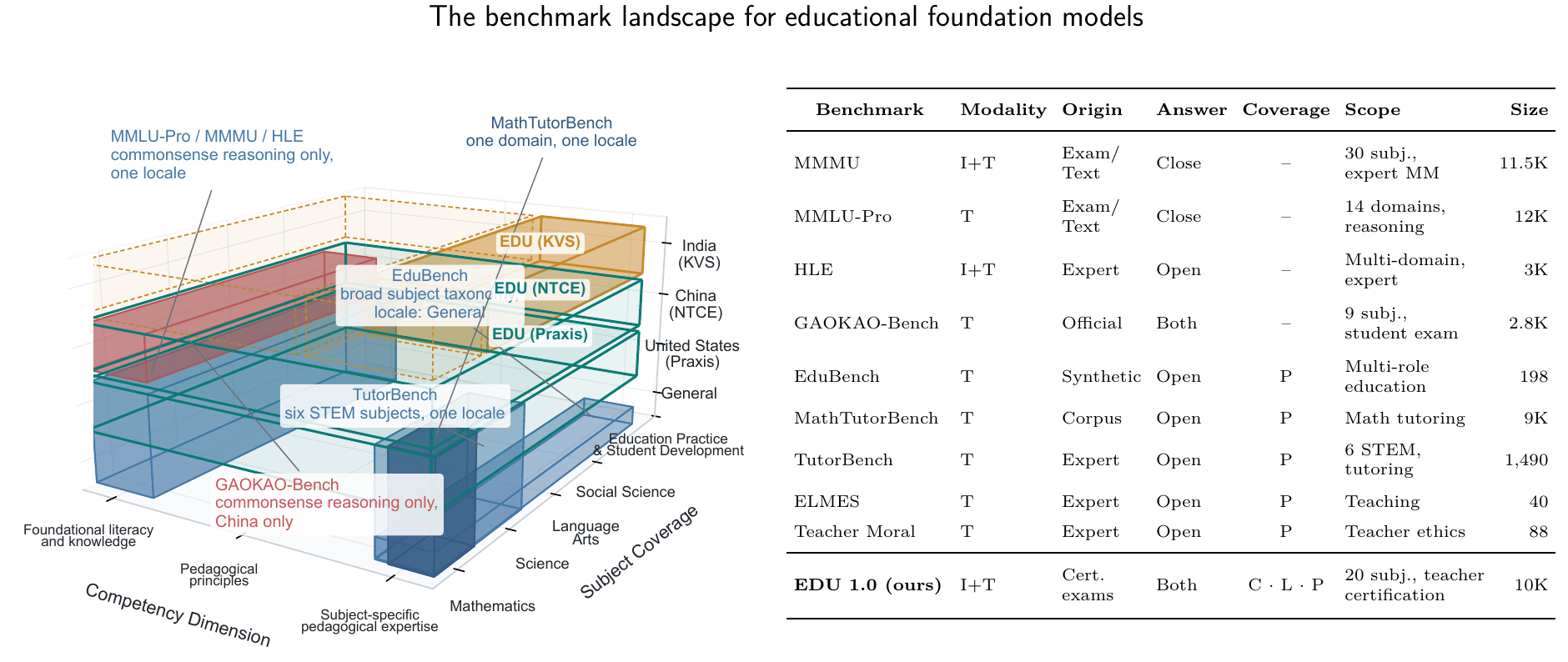}
    \caption{\textbf{Capability--domain--locale coverage of educational benchmarks.}
    \textbf{Left:} The three-dimensional evaluation space, with educational capability spanning commonsense reasoning, principle, and pedagogical content knowledge; educational domains spanning five broad subject groups; and educational locale spanning a General reference layer, U.S.\ Praxis, China NTCE and India KVS.
    Separate country-specific frames prevent the aggregate benchmark boundary from implying uniform coverage: U.S.\ Praxis and China NTCE span the full mapped grid, whereas solid ochre volumes mark KVS subject-specific coverage across four mapped domains and dashed ochre volumes denote regions not covered by KVS.
    \textbf{Right:} A systematic comparison of benchmark modality, origin, answer format, coverage, scope, and size; C = teacher-certification aligned, L = cross-locale, P = pedagogy, and \textit{Exam/Text} denotes examinations and textbooks.
    Existing benchmarks occupy restricted slices, whereas EDU 1.0 spans all three capability levels, 20 subjects, and three specified educational locales.
        }
    \label{fig:benchmark_space}
    \label{tab:comparison}
\end{figure}

To address this gap, we ask a concrete question against established professional standards: if foundation models are increasingly expected to support educational tasks traditionally performed by educators, do they exhibit the competencies represented in teacher-entry standards?
We introduce EDU 1.0 (\textit{Educational Due Diligence for Foundation Models})---a profession-aligned evaluation framework to inform evidence-based assessment before educational deployment~\cite{unesco2023genai}.
Teacher-entry examinations provide imperfect but externally established proxies for selected cognitive, pedagogical, and subject-specific dimensions of professional competence: they define what an entrant must know and how responses are judged~\cite{labue1960certification}. EDU 1.0 assembles these official examinations and scoring documents into a benchmark for foundation models.
The American case illustrates the institutional role of such standards. Certification emerged to protect instructional quality and the profession from unprepared entrants; over time, locally administered examinations of character and basic subject knowledge developed into state-administered standards encompassing academic skills, pedagogy, psychology, and discipline-specific knowledge~\cite{labue1960certification}. These dimensions are represented today in the Praxis Core Academic Skills, Principles of Learning and Teaching, and Subject Assessments~\cite{ets2024praxis}.

The institution is global: across 40 economies, nearly nine in ten require pedagogical training of prospective teachers, more than eight in ten a supervised practicum, and close to half a written examination~\cite{xie2026certification}. Systems that examine candidates in writing average 20 to 28 PISA points higher in reading, mathematics, and science, an association that provides suggestive evidence of the relevance of formal assessment systems but does not establish a causal effect~\cite{xie2026certification}.
EDU 1.0 is anchored in the Praxis Series of the United States, the National Teacher Certification Examination (NTCE) of China, and the Kendriya Vidyalaya Sangathan (KVS) teacher-recruitment examinations of India. Together they enrol more than 580 million students, employ 32 million teachers, and spend at the trillion scale annually (\Cref{fig:edu_infrastructure})~\cite{moe2025communique,nces2024digest,udise2025report,moe2025funding,census2025schoolfinances,indiabudget2026}. The three systems provide complementary coverage of a nationally coordinated East Asian certification system, a federal Western certification landscape, and a large multilingual developing education system---a range chosen in line with UNESCO's position that educational AI must serve developed and developing systems alike~\cite{miao2021policy}. We compare models across the cognitive, pedagogical, and subject-specific dimensions shared by these systems, without treating their distinct examinations as institutionally interchangeable.

\section{The Construction of EDU 1.0}\label{sec:data}
EDU 1.0 is constructed from official teacher-entry assessments and
assessment-aligned materials that operationalize selected professional
expectations for entry into teaching. Education experts and practising
teachers curated items from official examination archives and recognized
preparation materials.
Items were retained only when their original assessment
intent and scoring criteria could be verified. Each question was annotated
according to its source framework.
The annotation preserves the original assessment intent,
competency structure and scoring information, including the assessment module,
response format, reference answer and language of
administration~\cite{ets2024praxis,moe2013ntce}.
These assessments are treated as institutional indicators of selected
professional expectations rather than direct measures of classroom
effectiveness.

\begin{figure}[H]
    \centering
    \includegraphics[width=\textwidth]{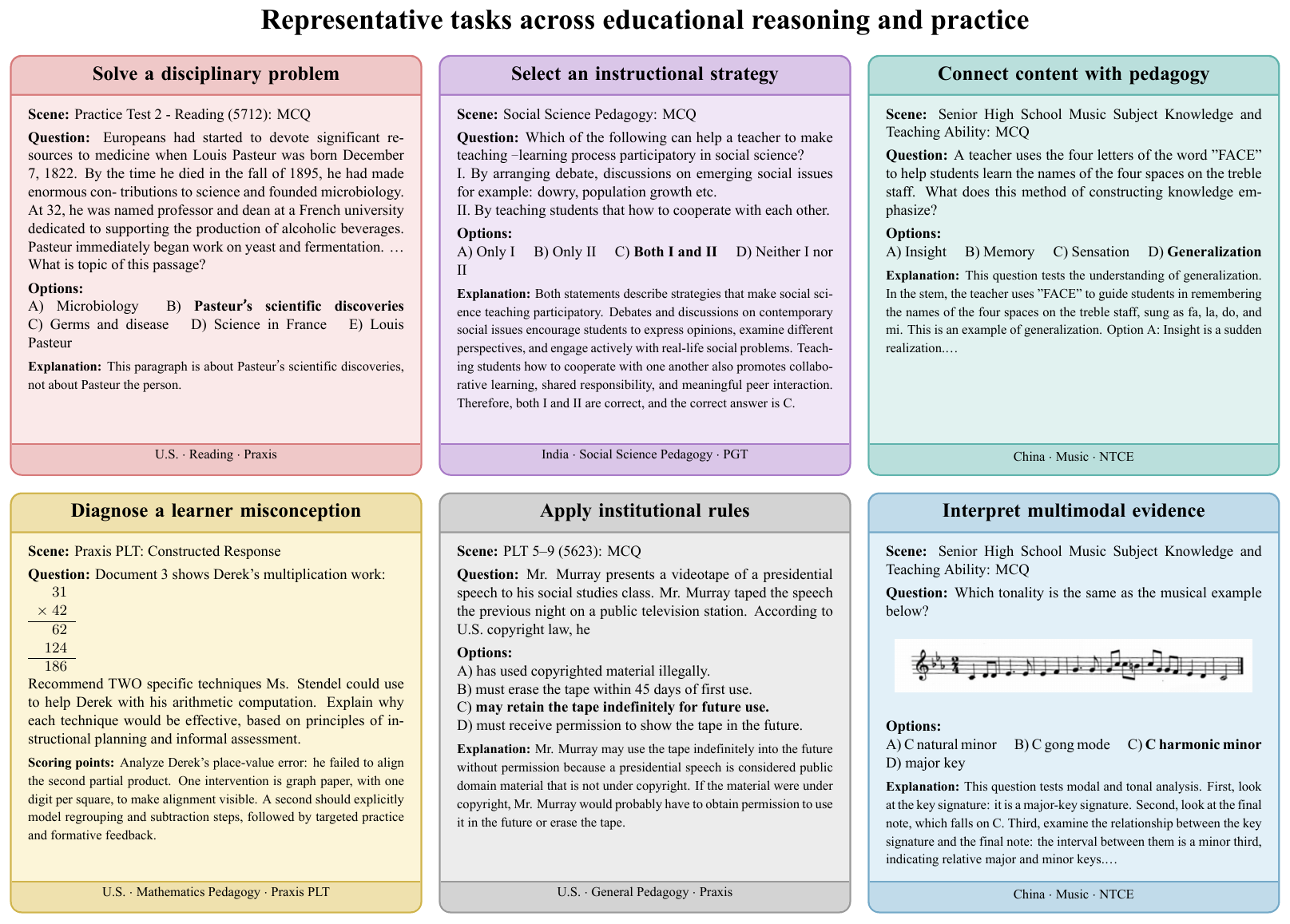}
    \caption{\textbf{Representative tasks across educational reasoning and practice.}
    Six example items from EDU 1.0, one per professional operation: solving a disciplinary problem, selecting an instructional strategy, connecting content with pedagogy, diagnosing a learner misconception, applying institutional rules, and interpreting multimodal evidence.
    Top labels indicate the professional operation required by each task, while footers identify its national system, disciplinary domain, and examination source.
    }
    \label{fig:competency_cards}
\end{figure}

\begin{figure}[H]
    \centering
    \includegraphics[width=\textwidth]{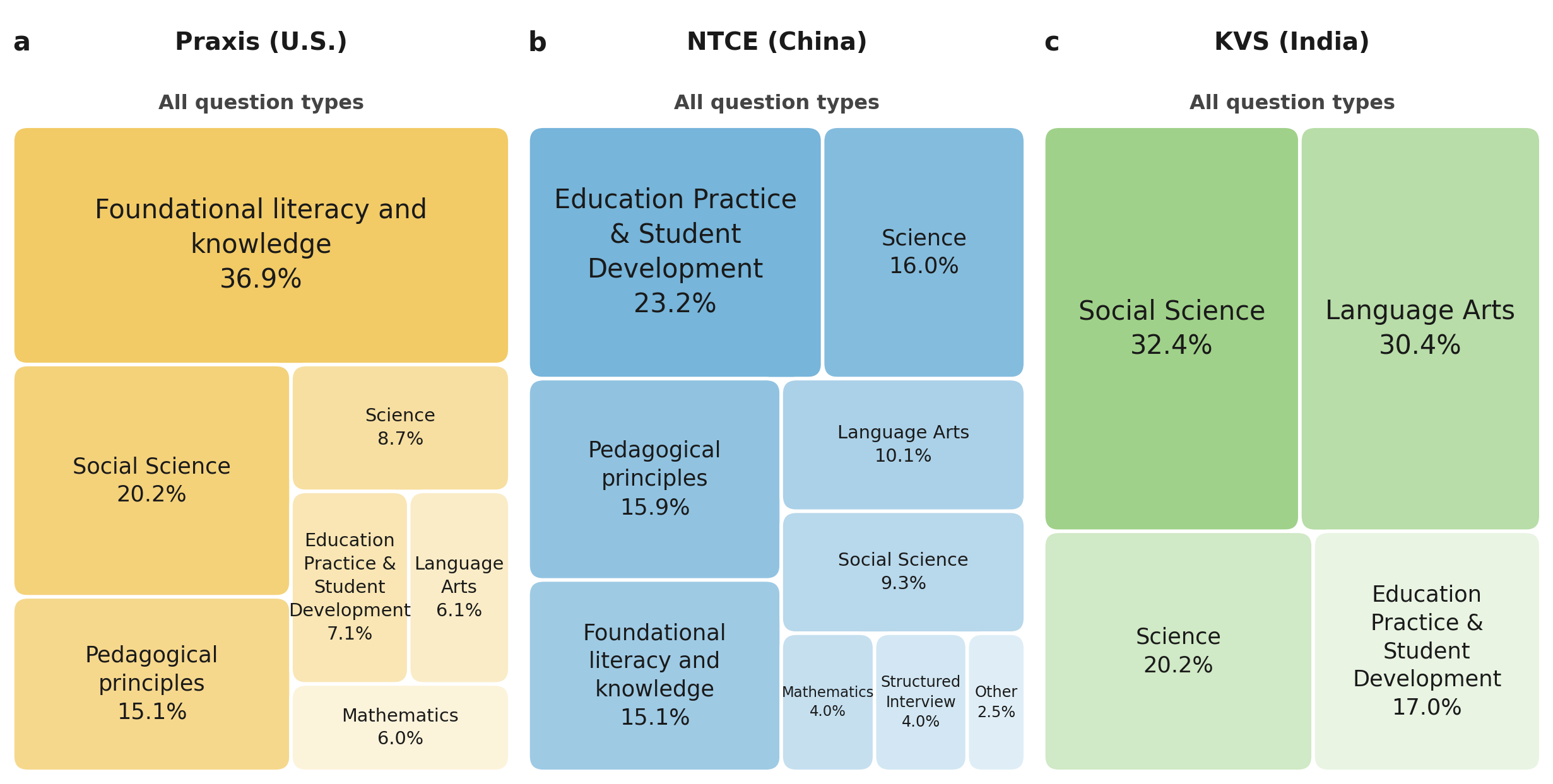}
    \caption{Composition of EDU 1.0 by national system. Each panel reports the share of all retained questions assigned to the aligned subject and competency categories within the U.S.\ Praxis, Chinese NTCE and Indian KVS collections. Percentages are normalized separately within each national system; the Indian collection contains selected-response questions only.}
    \label{fig:dataset_statistics}
\end{figure}

Despite differences in educational systems and cultural contexts, the
teacher-entry assessments represented in EDU 1.0 exhibit a degree of
structural convergence. They operationalize professional expectations through
three complementary dimensions:
foundational literacy and knowledge,
pedagogical principles,
and subject-specific pedagogical expertise~\cite{xie2026certification,shulman1986knowledge,ets2024praxis,moe2013ntce}.
For cross-system comparison, we further establish a subject-level taxonomy
spanning Language Arts, Mathematics, Science, Social Science, and Education
Practice \& Student Development. This taxonomy provides comparable analytical
categories; it does not replace the original disciplinary structures, which
remain available through the source-system annotations. The broad Education
Practice \& Student Development category collects professional and
learner-facing domains that lack direct disciplinary counterparts across all
three systems~\cite{shulman1986knowledge,intasc2024}.
\Cref{fig:dataset_statistics} summarizes the resulting composition of EDU 1.0
across the three national systems under this unified taxonomy, and
\Cref{fig:competency_cards} illustrates the range of professional operations
represented in the assessments, from disciplinary problem solving to the
diagnosis of learner misconceptions.

\subsection{The Praxis Assessment}
\label{sec:data:us}
The U.S.\ Praxis series represents three dimensions used in the cross-system
alignment: Core Academic Skills assesses foundational reasoning, Principles of
Learning and Teaching (PLT) assesses pedagogical knowledge and professional
judgement, and Subject Assessments address discipline-specific knowledge and
its application in teaching domains~\cite{ets2024praxis}. EDU 1.0 retains this
certification-defined structure while adding a hierarchical annotation scheme
for comparison with the Chinese and Indian systems.

EDU 1.0 captures this certification-defined competency spectrum through
1,903 retained questions---1,786 selected-response and 117 constructed-response
items---derived from official ETS Study Companions and publicly available
Praxis preparation materials. Each question was carefully aligned with the
original Praxis assessment framework and annotated using a hierarchical
two-level taxonomy. The first-level (L1) label preserves the certification
pathway defined by ETS: Praxis Core, Principles of Learning and Teaching (PLT),
and Subject Assessments. The second-level (L2) label further resolves the
underlying competency domain:
Core and PLT retain their pathway-level labels because they assess cross-cutting
academic and pedagogical competencies, whereas Subject Assessments are
consolidated into five disciplinary categories. Praxis English Language Arts
and Teaching Reading are mapped to Language Arts;
mathematics assessments across grade levels are mapped to Mathematics;
Biology, Physics and other natural-science assessments are mapped to Science;
and Citizenship and Social Studies assessments are mapped to Social Science.
Art, Health and Physical Education, School Counselor, and other professional
education assessments are grouped under Education Practice \& Student
Development.
This hierarchical
representation preserves the institutional origins of each assessment while
enabling evaluation of foundation models across a shared competency
space~\cite{ets2024praxis,shulman1986knowledge}.
The Praxis collection comprises 1{,}786 selected-response and 117
constructed-response items, a ratio inherited from the examinations themselves:
constructed responses appear only in Core Writing (two essays) and the PLT
assessments (four tasks each)~\cite{ets2024praxis}.
Pooling multiple editions of official and published preparation materials
therefore yields substantially more items of both kinds than a candidate
encounters in a single administration.

\subsection{The National Teacher Qualification Examination}
\label{sec:data:china}

To extend EDU 1.0 beyond the decentralized certification setting represented
by the United States and incorporate a nationally coordinated teacher-entry
system, we include China's National Teacher Qualification Examination (NTCE).
Unlike the decentralized
certification landscape in the United States, where requirements vary across
states and assessment providers, the NTCE operates under a nationally
coordinated framework that defines common
expectations for prospective teachers across regions.
This nationwide coordination, together with the scale of China's
teacher workforce, provides a large assessment resource collected under a
common framework. The NTCE subset therefore evaluates
model performance on selected professional expectations represented in a
large-scale national teacher-entry system~\cite{guo2024ntqe,liu2024reform,moe2025communique}.

EDU 1.0 retains the NTCE structure for secondary-school candidates: written
papers in Comprehensive Literacy (S1), Pedagogical Knowledge and Ability (S2),
and Subject Knowledge and Teaching Ability (S3), together with the structured
questions and teaching demonstration administered during the interview stage.
The written examination covers professional ethics and regulations, cultural
literacy and reasoning, education and classroom management, and
subject-specific knowledge and instructional design; the interview comprises
lesson preparation, structured questions, a teaching demonstration and
follow-up questioning~\cite{moe2013ntce}.
These interview-stage formats are retained as assessment artefacts and
evaluated only through their text-accessible components; they do not reproduce
the vocal, gestural or interactive conditions of an in-person performance.

EDU 1.0 captures this pathway through 4,688 retained questions---3,660
selected-response and 1,028 constructed-response items---drawn from NTCE past
examination papers and examination-aligned structured-interview and
teaching-demonstration collections~\cite{moe2013ntce,guo2024ntqe}.
Each question is aligned with the NTCE
framework and annotated using the same two-level taxonomy.
For comparative analysis, S1, S2 and S3 are respectively grouped with Praxis
Core, PLT and Subject
Assessments according to their dominant assessed dimensions; these mappings
identify shared dimensions rather than institutional equivalence. The NTCE
additionally contributes a Structured
Interview and Teaching Demonstration. At L2, S1 and S2
retain the labels Comprehensive Literacy and Knowledge and Ability of
Education.
The 13 S3 disciplines are grouped according to the dominant knowledge domain
assessed by each examination into the five domains used for Praxis Subject
Assessments---Language Arts, Mathematics, Science, Social Science, and
Education Practice \& Student Development---while their original discipline
labels are preserved. The two interview-stage formats remain distinct
categories~\cite{ets2024praxis,moe2013ntce}.
NTCE written examinations distribute their 150 raw points unevenly across
formats: selected-response items are worth two to five points each, whereas
material analysis, short answer, essay and instructional-design tasks are worth
eight to seventy, so a handful of constructed responses can carry more than half
of a module's score~\cite{moe2013ntce}. Our collection reflects this structure,
comprising 3{,}660 selected-response and 1{,}028 constructed-response items, and
is scored item by item on a normalized percentage scale rather than through the
official conversion.

\subsection{Kendriya Vidyalaya Sangathan Teacher Recruitment Examinations}
\label{sec:data:india}
Performance on foundational literacy or general pedagogical knowledge does not
necessarily predict performance on subject-specific teaching tasks. Our pilot
tests indicated a more persistent limitation on questions requiring
disciplinary knowledge to be applied in instructional contexts, a dimension
that overlaps with, but does not encompass, the broader construct of
pedagogical content knowledge. We therefore include Indian
recruitment examinations spanning post-specific teacher roles. The substantial
teacher and learner scale of India's education system, together with its
cultural and linguistic diversity, broadens the institutional and multilingual
coverage of EDU 1.0, including assessment items in English and
Hindi~\cite{shulman1986knowledge,udise2025report}.

We capture this combination using 3,421 retained multiple-choice questions
drawn from KVS Post Graduate Teacher (PGT), Educational and Vocational
Guidance Counsellor (EVGC), and related education-post
examinations.
Because recruitment is organized by professional post, these papers are
role-specific by construction.
They occupy a professional-entry stage comparable, for analytical purposes,
to the Subject Knowledge and Teaching Ability module (S3) of the Chinese NTCE
(\Cref{sec:data:china}) and the Praxis Subject Assessments
(\Cref{sec:data:us}); their second-level labels are therefore drawn from the
same analytical domains without treating the institutional pathways as
equivalent.
For the Indian collection, English and Hindi are mapped to Language Arts; chemistry, physics,
biology, mathematics and computer science to Science; history, economics and
commerce to Social Science; EVGC to Education Practice \& Student Development;
and physical education to Art \& Physical Education. As these examinations
contain selected-response questions only, no additional constructed-response
or interview-stage category is introduced.

\section{Evaluation} \label{sec:evaluation}

\subsection{Setup}
EDU 1.0 spans question formats whose scoring demands differ fundamentally. We therefore separate the benchmark into two regimes standard in educational measurement: selected-response questions, which have a fixed answer key, and constructed-response tasks---written subjective answers, interviews, and teaching demonstrations---which admit multiple valid responses.
All scoring is performed at the item level before aggregation, preserving the original within-item scoring structure while enabling normalization across heterogeneous scales.

Selected-response items comprise two formats: single-choice questions with exactly one correct option, and multiple-select questions, such as the Praxis \emph{select all that apply} items, whose answer key contains several options.
Both are scored using rule-based answer extraction followed by exact comparison with the official answer key.
After removing explicit reasoning tags, we extract option strings from free-form outputs in a fixed order: an explicit answer statement, a stand-alone answer on the final line, or the final stand-alone option token; option sets written contiguously (``AB'') or with separators (``A, C'') are normalized to a common form.
These extraction rules were fixed before evaluation and applied identically across all models. A single-choice response is correct when the extracted option matches the reference option, whereas a multiple-select response is correct only when the extracted set equals the complete reference set, irrespective of order, so that partially correct selections receive no credit.
Unparseable outputs are marked incorrect and no partial credit or LLM-assisted fallback is used; this conservative policy avoids introducing evaluator-dependent variability into selected-response scoring.
For model $m$ on item $i$, let $\hat{y}_{mi}$ denote the set of options extracted from its response by the rules above, taken to be the empty set when no option can be extracted, and let $y_i$ denote the official answer key, a singleton for single-choice items and a set of several options for multiple-select items. The item result is
\begin{equation}
    r_{m,i}=100\cdot\mathbb{I}\left(\hat{y}_{m,i}=y_i\right)
    \label{eq:selected-response-item-result}
\end{equation}
so that set equality expresses both scoring rules at once: for single-choice items it reduces to matching the reference option, and for multiple-select items it requires the complete reference set.

Scoring constructed responses requires judgement against a rubric rather than a
lookup against a key. To balance between scalability and reliability,
we delegate that judgement to a fixed panel of language models against the
official rubrics and reference answer to every response.
To keep the scoring reproducible from released checkpoints, we draw the panel
exclusively from open-weight systems, and within that scope we include every
system in the leading performance tier---Kimi-K25 and Qwen3.5-397B---which score
each response independently and, coming from different model families, keep the
panel from resting on a single lineage.
Judge-level scores are then averaged for each model--item pair, so that each item contributes one aggregated score per model. Let $\mathcal{J}_{mi}$ denote the judges that return a valid score for model $m$ on item $i$; for judge $j$, let $s_{mij}$ be the raw score it awards and $M_{mij}$ the maximum score available for that item under its rubric. We~compute the per-judge percentage, the mean raw score and the item result,
\begin{equation}
    {p}_{mij}=100\,\frac{s_{mij}}{M_{mij}},\qquad
    r_{mi}=\frac{1}{|\mathcal{J}_{mi}|}\sum_{j\in\mathcal{J}_{mi}}{p}_{mij}.
    \label{eq:constructed-response-score}
\end{equation}
Averaging in this form preserves the official within-item scale in $\bar{s}_{mi}$, while $r_{mi}$ is comparable across items whose maxima differ. Items with partial model coverage are omitted from shared comparison tables by default.
Every evaluated item, of either format, therefore carries a single result $r_{mi}\in[0,100]$ on a common percentage scale: $0$ or $100$ for a selected response according to its correctness, and the awarded percentage score for a constructed one. The analyses that follow differ only in how they aggregate $r_{mi}$ over items. Written tasks use type-specific certification rubrics, and structured interviews use the four text-assessable dimensions of the NTCE rubric (40 points).
Teaching demonstrations are assessed for lesson objectives, instructional organization, explanation quality and learner-oriented adaptation.
Three education experts independently scored 500 constructed responses across 14
evaluation subsets, agreeing closely with one another (Krippendorff's
$\alpha=0.898$). The two-judge panel reproduces their scores more faithfully than
either judge alone (Spearman $\rho=0.864$, quadratic weighted $\kappa=0.783$).
Scoring rubrics, extraction rules, aggregation details and judge prompts are
provided in the released source code.

\paragraph{Models compared.}
We evaluate 36 model configurations. Thinking and non-thinking modes of the same base model are treated as distinct configurations and are not pooled: complete roster is provided in
\Cref{tab:teach_cross_national_ci}.
Whether capabilities are practically accessible to education systems operating under different resource constraints (\Cref{sec:introduction}) depends on how a model can be obtained and run, and we group the evaluated configurations along this dimension rather than by capability: T1 comprises proprietary systems available only through hosted APIs; T2 comprises weight-released systems that require industrial-scale, typically multi-accelerator, inference infrastructure, above 200B total parameters in this roster; and T3 comprises weight-released systems amenable to local inference using quantization or inference acceleration, at or below 50B.
Releasing weights does not by itself confer access, which also depends on memory, compute, serving infrastructure and deployment expertise; these groups are therefore operational deployment categories rather than measures of accessibility.

\subsection{Results on Cross-National Constructed-Response and Selected-Response Questions}
\label{sec:results:cross-national}

\begin{table}[!t]\small
    \centering
    \renewcommand\arraystretch{1.05}
    \renewcommand\tabcolsep{6pt}
    \caption{Cross-national teacher-certification results with bootstrap confidence intervals. Each cell reports the mean score followed by its 95\% percentile-bootstrap interval over 10,000 resamples of the underlying questions, computed as in \Cref{eq:cross-national-ci} with a fixed random seed so that every interval is reproducible. Selected- and constructed-response items are pooled within each system into a single country score. Item-average weights every evaluated item equally; response-balanced weights the five country-by-response-type strata at 1/6, 1/6, 1/6, 1/6 and 1/3 so that neither response type nor the larger Indian item pool dominates. Two cases are treated exactly rather than by simulation: where a subset is scored purely right or wrong, resampling and counting the correct answers is distributed exactly as a binomial, which we sample directly; and where a model answers every item of a subset correctly or incorrectly the percentile bootstrap degenerates to a point, so we report the exact Clopper--Pearson interval instead. Intervals cover sampling variability over items only\,---\,constructed-response scores are means over three judges, and that variation is averaged into the item score before resampling. Within each access tier, models with \textit{top-3} scores in each column are marked with best in \textbf{bold} and second/third \underline{underlined}. Intervals are comparable within a column only: different columns rest on different item pools. Inference Cost is the OpenRouter list price in USD per million output tokens.}
    \vspace{4pt}
    \resizebox{\linewidth}{!}{\begin{tabular}{l|ccccc|c}
    \toprule
        \multirow{2}{*}{\textbf{Model}} & \textbf{Praxis (U.S.)} & \textbf{NTCE (China)} & \textbf{KVS (India)} & \textbf{Item-average} & \textbf{Response-balanced} & \textbf{Inference Cost} \\
        \cmidrule(lr){2-2} \cmidrule(lr){3-3} \cmidrule(lr){4-4} \cmidrule(lr){5-5} \cmidrule(lr){6-6} \cmidrule(lr){7-7}
        & \makecell{\textit{Score}\\\textit{(95\% CI)$\uparrow$}} & \makecell{\textit{Score}\\\textit{(95\% CI)$\uparrow$}} & \makecell{\textit{Score}\\\textit{(95\% CI)$\uparrow$}} & \makecell{\textit{Score}\\\textit{(95\% CI)$\uparrow$}} & \makecell{\textit{Score}\\\textit{(95\% CI)$\uparrow$}} & \makecell{\textit{USD /}\\\textit{M output tokens$\downarrow$}} \\
    \midrule
    \multicolumn{7}{l}{\textit{Tier 1 (T1): proprietary API-only models}} \\
    \midrule
    Claude Opus 4.6 & \underline{93.6}~{\scriptsize [92.5, 94.6]} & \underline{93.2}~{\scriptsize [92.5, 93.8]} & \underline{90.6}~{\scriptsize [89.6, 91.6]} & \underline{92.4}~{\scriptsize [91.9, 92.8]} & \underline{93.4}~{\scriptsize [92.9, 93.8]} & \$25 \\
    Claude Sonnet 4.6 & 91.7~{\scriptsize [90.5, 92.9]} & 90.6~{\scriptsize [89.8, 91.3]} & 89.9~{\scriptsize [88.8, 90.9]} & 90.6~{\scriptsize [90.0, 91.1]} & 91.8~{\scriptsize [91.3, 92.3]} & \$15 \\
    Gemini 3.1 Pro & \textbf{97.7}~{\scriptsize [96.9, 98.3]} & \textbf{95.2}~{\scriptsize [94.7, 95.7]} & \textbf{97.2}~{\scriptsize [96.6, 97.7]} & \textbf{96.3}~{\scriptsize [96.0, 96.6]} & \textbf{96.5}~{\scriptsize [96.2, 96.8]} & \$12 \\
    Gemini 3 Flash & \underline{95.0}~{\scriptsize [94.0, 95.9]} & \underline{93.9}~{\scriptsize [93.3, 94.5]} & \underline{95.8}~{\scriptsize [95.1, 96.5]} & \underline{94.8}~{\scriptsize [94.4, 95.2]} & \underline{95.0}~{\scriptsize [94.6, 95.3]} & \$3 \\
    GPT 5.2 & 86.1~{\scriptsize [84.5, 87.6]} & 88.1~{\scriptsize [87.2, 88.9]} & 80.2~{\scriptsize [78.9, 81.5]} & 85.0~{\scriptsize [84.3, 85.7]} & 87.2~{\scriptsize [86.6, 87.8]} & \$14 \\
    \midrule
    \multicolumn{7}{l}{\textit{Tier 2 (T2): industrial-scale open-weight models}} \\
    \midrule
    Kimi K2.5 & \underline{93.7}~{\scriptsize [92.6, 94.7]} & \underline{93.4}~{\scriptsize [92.8, 94.0]} & \underline{89.9}~{\scriptsize [88.9, 90.8]} & \underline{92.2}~{\scriptsize [91.7, 92.8]} & \underline{93.0}~{\scriptsize [92.6, 93.5]} & \$2.025 \\
    Qwen3.5 397B & \textbf{95.1}~{\scriptsize [94.2, 96.1]} & \textbf{94.7}~{\scriptsize [94.2, 95.2]} & \textbf{92.6}~{\scriptsize [91.8, 93.5]} & \textbf{94.1}~{\scriptsize [93.6, 94.5]} & \textbf{94.4}~{\scriptsize [94.0, 94.8]} & \$2.45 \\
    qwen3.5 397b-No Think & \underline{92.0}~{\scriptsize [90.8, 93.2]} & \underline{93.3}~{\scriptsize [92.7, 93.9]} & \underline{85.4}~{\scriptsize [84.3, 86.6]} & \underline{90.4}~{\scriptsize [89.8, 90.9]} & \underline{91.2}~{\scriptsize [90.6, 91.7]} & \$2.45 \\
    GLM-4.6V & 90.7~{\scriptsize [89.4, 91.9]} & 91.1~{\scriptsize [90.4, 91.8]} & 82.9~{\scriptsize [81.6, 84.1]} & 88.2~{\scriptsize [87.6, 88.8]} & 88.6~{\scriptsize [88.0, 89.2]} & \$0.9 \\
    \midrule
    \multicolumn{7}{l}{\textit{Tier 3 (T3): locally deployable open-weight models}} \\
    \midrule
    Gemma 4 12B IT-No Think & 82.8~{\scriptsize [81.0, 84.4]} & 76.3~{\scriptsize [75.2, 77.4]} & 70.9~{\scriptsize [69.4, 72.5]} & 75.7~{\scriptsize [74.9, 76.5]} & 79.0~{\scriptsize [78.3, 79.8]} & Local\textsuperscript{*} \\
    Gemma 4 26B-A4B IT-No Think & 81.5~{\scriptsize [79.7, 83.2]} & 78.4~{\scriptsize [77.3, 79.5]} & 77.9~{\scriptsize [76.6, 79.3]} & 78.8~{\scriptsize [78.0, 79.6]} & 77.5~{\scriptsize [76.8, 78.3]} & \$0.34 \\
    Gemma 4 31B IT-No Think & \underline{90.9}~{\scriptsize [89.5, 92.1]} & 84.6~{\scriptsize [83.7, 85.5]} & \underline{82.5}~{\scriptsize [81.2, 83.8]} & 85.1~{\scriptsize [84.4, 85.7]} & 86.7~{\scriptsize [86.1, 87.3]} & \$0.34 \\
    Gemma 4 E2B IT-No Think & 66.4~{\scriptsize [64.3, 68.5]} & 61.0~{\scriptsize [59.7, 62.3]} & 56.2~{\scriptsize [54.6, 57.9]} & 60.4~{\scriptsize [59.5, 61.3]} & 66.2~{\scriptsize [65.3, 67.0]} & Local\textsuperscript{*} \\
    InternVL3.5 14B-No Think & 83.4~{\scriptsize [81.8, 85.1]} & 81.7~{\scriptsize [80.8, 82.7]} & 70.2~{\scriptsize [68.6, 71.7]} & 78.1~{\scriptsize [77.3, 78.9]} & 78.2~{\scriptsize [77.4, 78.9]} & Local\textsuperscript{*} \\
    InternVL3.5 2B-No Think & 65.5~{\scriptsize [63.4, 67.6]} & 64.8~{\scriptsize [63.5, 66.0]} & 54.7~{\scriptsize [53.0, 56.4]} & 61.5~{\scriptsize [60.5, 62.4]} & 60.5~{\scriptsize [59.7, 61.4]} & Local\textsuperscript{*} \\
    InternVL3.5 4B-No Think & 75.1~{\scriptsize [73.2, 76.9]} & 75.5~{\scriptsize [74.4, 76.5]} & 61.9~{\scriptsize [60.3, 63.5]} & 70.7~{\scriptsize [69.9, 71.6]} & 70.4~{\scriptsize [69.6, 71.2]} & Local\textsuperscript{*} \\
    InternVL3.5 8B-No Think & 77.8~{\scriptsize [76.0, 79.6]} & 78.9~{\scriptsize [77.9, 79.9]} & 67.0~{\scriptsize [65.3, 68.5]} & 74.6~{\scriptsize [73.8, 75.4]} & 74.5~{\scriptsize [73.7, 75.2]} & Local\textsuperscript{*} \\
    Kimi-VL A3B Instruct-No Think & 72.3~{\scriptsize [70.3, 74.2]} & 77.3~{\scriptsize [76.3, 78.3]} & 64.1~{\scriptsize [62.5, 65.7]} & 71.8~{\scriptsize [71.0, 72.6]} & 70.5~{\scriptsize [69.7, 71.3]} & Local\textsuperscript{*} \\
    MiMo-VL 7B SFT-No Think & 83.9~{\scriptsize [82.2, 85.5]} & 80.7~{\scriptsize [79.7, 81.7]} & 71.1~{\scriptsize [69.6, 72.6]} & 78.0~{\scriptsize [77.2, 78.8]} & 79.4~{\scriptsize [78.7, 80.1]} & Local\textsuperscript{*} \\
    Ministral 3 14B & 79.1~{\scriptsize [77.2, 80.9]} & 78.3~{\scriptsize [77.2, 79.4]} & 69.3~{\scriptsize [67.7, 70.8]} & 75.4~{\scriptsize [74.6, 76.2]} & 78.4~{\scriptsize [77.7, 79.1]} & \$0.2 \\
    Ministral 3 3B & 70.4~{\scriptsize [68.4, 72.5]} & 66.5~{\scriptsize [65.2, 67.7]} & 59.8~{\scriptsize [58.2, 61.4]} & 64.9~{\scriptsize [64.0, 65.8]} & 68.7~{\scriptsize [67.9, 69.5]} & \$0.1 \\
    Ministral 3 8B & 78.0~{\scriptsize [76.1, 79.9]} & 77.2~{\scriptsize [76.2, 78.3]} & 67.4~{\scriptsize [65.9, 69.0]} & 74.0~{\scriptsize [73.2, 74.8]} & 77.0~{\scriptsize [76.3, 77.8]} & \$0.15 \\
    Molmo2 8B-No Think & 76.4~{\scriptsize [74.5, 78.3]} & 81.0~{\scriptsize [80.0, 82.0]} & 67.5~{\scriptsize [65.9, 69.1]} & 75.5~{\scriptsize [74.7, 76.3]} & 76.0~{\scriptsize [75.2, 76.8]} & Local\textsuperscript{*} \\
    Phi-3.5 Vision & 61.9~{\scriptsize [59.8, 64.1]} & 44.5~{\scriptsize [43.2, 45.8]} & 57.1~{\scriptsize [55.5, 58.8]} & 52.1~{\scriptsize [51.2, 53.1]} & 54.7~{\scriptsize [53.9, 55.6]} & Local\textsuperscript{*} \\
    Qwen3.5 4B & 89.8~{\scriptsize [88.4, 91.1]} & 87.0~{\scriptsize [86.1, 87.8]} & 76.3~{\scriptsize [74.9, 77.7]} & 83.9~{\scriptsize [83.2, 84.5]} & 84.6~{\scriptsize [84.0, 85.3]} & Local\textsuperscript{*} \\
    Qwen3.5 4B-No Think & 78.2~{\scriptsize [76.3, 80.1]} & 83.9~{\scriptsize [82.9, 84.8]} & 65.0~{\scriptsize [63.3, 66.5]} & 76.3~{\scriptsize [75.5, 77.1]} & 78.3~{\scriptsize [77.5, 79.0]} & Local\textsuperscript{*} \\
    Qwen3.5 9B & 88.2~{\scriptsize [86.8, 89.6]} & 90.6~{\scriptsize [89.8, 91.3]} & 81.6~{\scriptsize [80.3, 82.9]} & \underline{87.1}~{\scriptsize [86.4, 87.7]} & \underline{87.9}~{\scriptsize [87.4, 88.5]} & \$0.15 \\
    Qwen3.5 9B-No Think & 80.0~{\scriptsize [78.2, 81.8]} & 87.3~{\scriptsize [86.5, 88.2]} & 69.5~{\scriptsize [67.9, 71.0]} & 79.8~{\scriptsize [79.1, 80.6]} & 81.8~{\scriptsize [81.1, 82.5]} & \$0.15 \\
    Qwen3.6 27B & \textbf{95.7}~{\scriptsize [94.7, 96.5]} & \textbf{92.9}~{\scriptsize [92.2, 93.5]} & \underline{87.0}~{\scriptsize [85.9, 88.1]} & \textbf{91.4}~{\scriptsize [90.9, 91.9]} & \textbf{92.2}~{\scriptsize [91.7, 92.6]} & \$2.4 \\
    Qwen3.6 27B-No Think & 86.9~{\scriptsize [85.4, 88.4]} & \underline{91.6}~{\scriptsize [90.9, 92.3]} & 78.4~{\scriptsize [77.0, 79.8]} & 86.2~{\scriptsize [85.5, 86.8]} & 87.4~{\scriptsize [86.8, 88.0]} & \$2.4 \\
    Qwen3.5 35B & \underline{91.3}~{\scriptsize [90.0, 92.6]} & \underline{92.5}~{\scriptsize [91.9, 93.2]} & \textbf{88.1}~{\scriptsize [87.1, 89.2]} & \underline{90.8}~{\scriptsize [90.2, 91.3]} & \underline{91.4}~{\scriptsize [90.9, 91.9]} & \$1 \\
    Qwen3.5 35B-No Think & 84.1~{\scriptsize [82.5, 85.8]} & 90.7~{\scriptsize [90.0, 91.5]} & 76.9~{\scriptsize [75.5, 78.3]} & 84.8~{\scriptsize [84.1, 85.4]} & 86.2~{\scriptsize [85.5, 86.8]} & \$1 \\
    Qwen3.6 35B & 86.6~{\scriptsize [85.1, 88.2]} & 90.8~{\scriptsize [90.0, 91.5]} & 76.6~{\scriptsize [75.2, 78.0]} & 85.2~{\scriptsize [84.5, 85.8]} & 86.4~{\scriptsize [85.8, 87.0]} & \$1 \\
    Qwen3.6 35B-No Think & 85.9~{\scriptsize [84.3, 87.4]} & 90.7~{\scriptsize [90.0, 91.4]} & 77.1~{\scriptsize [75.7, 78.5]} & 85.1~{\scriptsize [84.5, 85.8]} & 86.4~{\scriptsize [85.8, 87.1]} & \$1 \\
    Step3-VL 10B-No Think & 86.8~{\scriptsize [85.3, 88.3]} & 80.4~{\scriptsize [79.3, 81.4]} & 71.9~{\scriptsize [70.4, 73.4]} & 78.7~{\scriptsize [77.9, 79.5]} & 82.0~{\scriptsize [81.3, 82.7]} & Local\textsuperscript{*} \\
    Llama 3.2 11B Vision Instruct-No Think & 65.8~{\scriptsize [63.7, 67.9]} & 54.4~{\scriptsize [53.1, 55.6]} & 58.8~{\scriptsize [57.2, 60.5]} & 58.0~{\scriptsize [57.1, 59.0]} & 61.7~{\scriptsize [60.9, 62.6]} & Local\textsuperscript{*} \\
    \bottomrule
    \end{tabular}}
    \par\vspace{2pt}
    \parbox{\linewidth}{\tiny *~\textit{Local} indicates self-hosted open-weight models without API pricing, whose cost is set by GPU utilisation rather than a per-token tariff; the column is therefore not a homogeneous quantity and carries no top-3 marking.}
    \label{tab:teach_cross_national_ci}
\end{table}

\paragraph{Cross-national comparison.}
We first score every model within each national examination separately, so that no examination is averaged away before it can be inspected.
\Cref{tab:teach_cross_national_ci} reports one pooled score per system with its bootstrap interval, alongside the two cross-national aggregates.
    \Cref{tab:teach_cross_national_ci}
reports one pooled score per system with its 95\% bootstrap interval for all three systems: the  \Cref{eq:cross-national-aggregation}.

Comparing models across systems requires a single number, and we report two aggregate choice of aggregatio together with bootstrap confidence intervals to quantify statistical uncertainty.
Let $r_{mi}$ be the item result of model $m$ on item $i$, as defined in
\Cref{sec:evaluation}; let $\mathcal{C}$ denote the three national systems and
$\mathcal{T}_c$ the response types present in system $c$; let $c(i)$ and $t(i)$
be the system and response type that item $i$ belongs to; and let $N_{ct}$ be the
number of evaluated items of type $t$ in system $c$, with
$N=\sum_{c}\sum_{t}N_{ct}$. The two aggregates are
\begin{equation}
    \mathrm{IA}_m=\frac{1}{N}\sum_{i}r_{mi},
    \qquad
    \mathrm{CM}_m=\frac{1}{|\mathcal{C}|}\sum_{c\in\mathcal{C}}
    \frac{1}{|\mathcal{T}_c|}\sum_{t\in\mathcal{T}_c}
    \frac{1}{N_{ct}}\sum_{i:\,c(i)=c,\,t(i)=t} r_{mi}.
    \label{eq:cross-national-aggregation}
\end{equation}
Both are weighted item means $A_m=\sum_i w_i\,r_{mi}$, with
\begin{equation}
    w_i^{\mathrm{IA}}=\frac{1}{N},
    \qquad
    w_i^{\mathrm{CM}}=\frac{1}{|\mathcal{C}|\;|\mathcal{T}_{c(i)}|\;N_{c(i)t(i)}},
    \label{eq:cross-national-item-weights}
\end{equation}
so that $\mathrm{CM}$ fixes each system's share at $1/|\mathcal{C}|$ and each
response type's share within a system at $1/|\mathcal{T}_c|$, independently of
subset size.

\begin{samepage}
We attach a confidence interval to each aggregate by resampling \emph{items}, so
that an interval answers how much a score would move had the benchmark drawn a
different sample of questions from the same examinations. For $b=1,\dots,B$, let
$i_j^{(b)}\in\{1,\dots,N\}$ denote the original-item index on draw $j$ of replicate
$b$, sampled independently with replacement so that $\Pr(i_j^{(b)}=i)=w_i$. We
then recompute the aggregate on the resample:
\begin{equation}
    A^{(b)}_m=\frac{1}{N}\sum_{j=1}^{N}r_{m\,i^{(b)}_j},
    \qquad
    \mathrm{CI}_{1-\alpha}(A_m)=
    \Bigl[\,Q_{\alpha/2}\bigl(\{A^{(b)}_m\}\bigr),\;
            Q_{1-\alpha/2}\bigl(\{A^{(b)}_m\}\bigr)\,\Bigr],
    \label{eq:cross-national-ci}
\end{equation}
\end{samepage}
with $Q_q$ the $q$-th quantile over the $B$ resampled aggregates and
$\alpha=0.05$. Because $w_i^{\mathrm{IA}}$ is uniform, this reduces to the
ordinary percentile bootstrap for the item average; for the country macro it
draws items in proportion to their macro weight, so a small subset carrying a
large target share widens the interval rather than silently dominating the point
estimate.

\paragraph{Cross-national comparison.}
Across the full roster, Praxis separates weaker models---scores span
61.9\%--97.7\%---but compresses at the top, where the leader reaches
\ci{97.7}{96.9}{98.3} and the leader plus the next four configurations fall within 4.0 points of one
another. The NTCE and KVS examinations sit further from ceiling (95.2\% and
97.2\% at the top) and spread the roster more widely, across 50.6 and 42.5 points.

\paragraph{Leadership at two levels.}
We next ask who leads, first within each examination and then within each access tier.
At the examination level, leadership does not follow a model's origin:
the same ordering recurs across all three systems: model
 rankings correlate at $\rho=0.91$--$0.97$ between any pair of examinations (Spearman, $p<10^{-13}$)
, even though the systems differ in language, curriculum and response format.
Gemini 3.1 Pro leads by margins its intervals resolve on Praxis,
\cis{97.7}{96.9}{98.3}, and on the KVS papers, \cis{97.2}{96.6}{97.7}; on the
NTCE it is first at \cis{95.2}{94.7}{95.7} but not separably from Qwen3.5 397B.
No leading model is
specialized for the examinations it tops
, evidence of transfer across
institutionally and culturally distinct teacher-entry examinations.
At the tier level, both aggregates return the same ordering, and the intervals
separate consecutive tiers: the leading proprietary system reaches
\cis{96.3}{96.0}{96.6}, the leading industrial-scale open-weight system
\cis{94.1}{93.6}{94.5} and the leading locally deployable one
\cis{91.4}{90.9}{91.9} under equal item
weighting, with the same ordering under the response-balanced macro. Within T3,
however, the top two configurations overlap. The 2.2 and 4.9 points separating the T1 leader from the T2 and T3 leaders remain an
order of magnitude below the 44.2 points spanned by the roster; we return to what
this implies for deployment in \Cref{sec:discussion}.

 \subsection{Effect of Model Scaling}

\begin{figure}[H]
    \centering
        \includegraphics[width=\linewidth]{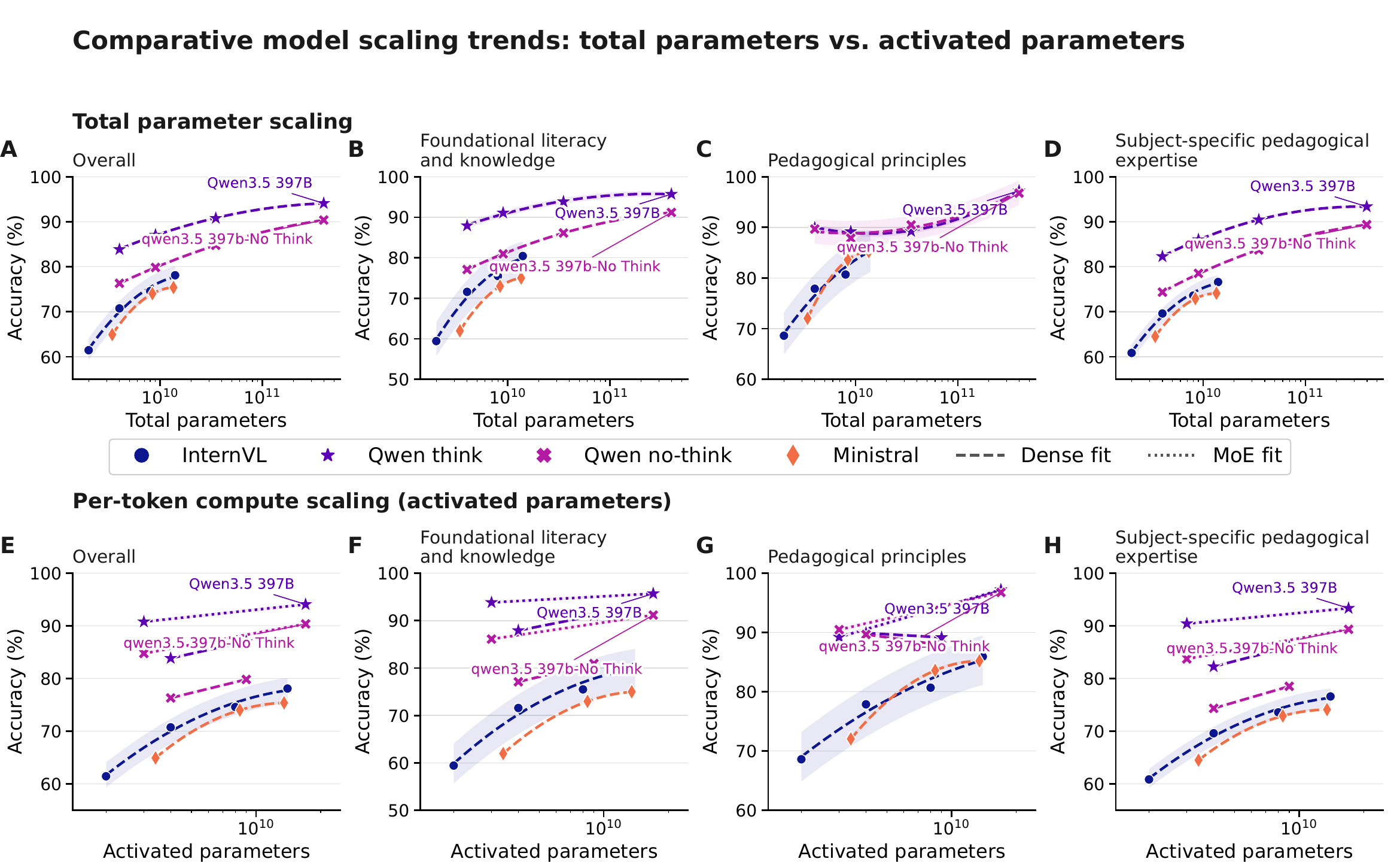}
    \caption{\textbf{Model scaling trends on EDU 1.0.}
    \textbf{A}--\textbf{D}, total language-model parameters; \textbf{E}--\textbf{H},
    activated language-model parameters. Both parameter axes are logarithmic.
    For each fitted group of \(n\) checkpoints, we set
    \(z_i=\log_{10}x_i\) and fit accuracy by least squares as \(p_d(z)\), where
    \(d=\min(2,n-1)\). With residuals \(e_i=y_i-p_d(z_i)\), shading shows
    \(p_d(z)\pm h\), where \(h=\max(1.96s,0.8)\) percentage points and
    \(s=\sqrt{\sum_i e_i^2/(n-d-1)}\) when \(n>d+1\); otherwise, \(s\) is the
    population standard deviation of the residuals. These fixed-width,
    residual-based envelopes are not formal confidence intervals.}
            \label{fig:parameter-scaling-row}
\end{figure}

To examine whether model scaling differs across the professional demands represented
in EDU 1.0, we break down performance using the three cross-system L1 categories
defined in \Cref{sec:data}: foundational literacy and knowledge, pedagogical
principles, and subject-specific pedagogical expertise.
The breakdowns reported in this and the following sections all take the same form,
so we fix the notation once. Let \(\gamma:\mathcal{I}\rightarrow\mathcal{G}\) be a
grouping rule assigning each evaluated item to a group, and let
\(\mathcal{I}_{mg}=\{i:\gamma(i)=g \}\) collect the
items of group \(g\) on which model \(m\) has a valid item result \(r_{mi}\)
(\Cref{sec:evaluation}). The group score is the unweighted mean
\begin{equation}
    A_{mg}
    =
    \frac{1}{|\mathcal{I}_{mg}|}
    \sum_{i\in\mathcal{I}_{mg}}r_{mi}.
    \label{eq:group-score}
\end{equation}
Here \(\gamma\) assigns each item to its L1 category. This aggregation gives each
evaluated item equal weight while retaining partial credit for constructed
responses, and pools examination modules only after they have been aligned to the
same L1 category. Each plotted point is one such aggregate score rather than a
mean over repeated model runs.

\paragraph{Summary of parameter scaling}

Panels \textbf{A}--\textbf{D} of \Cref{fig:parameter-scaling-row} compare
total-parameter scaling across four model families with known parameters---InternVL, Qwen 3.5 think,
Qwen 3.5 no-think, and Ministral---using multiple checkpoints per family. Each
panel reports either overall performance or one of the three aligned L1
categories: foundational literacy and knowledge, pedagogical principles, and
subject-specific pedagogical expertise.
Overall performance rises steadily with parameter count, but with diminishing
returns: successive Qwen checkpoints return 9.1, 6.3 and 3.1 points per decade of
parameters. Foundational literacy and subject-specific pedagogical expertise both
follow this pattern, improving monotonically with scale and benefiting from
extended reasoning by a margin that narrows as capacity grows, from $+10.8$ to
$+4.5$ points and from $+7.9$ to $+4.0$ points respectively. Pedagogical
principles behaves differently on both counts: it neither improves steadily
across the intermediate scales nor responds to extended reasoning at any of them,
the thinking margin staying between $-1.3$ and $+1.3$ points throughout, so this
competence is not one that scale alone appears to resolve.

\paragraph{Per-token FLOPS scaling.}
Serving cost is governed by the computation activated per token rather than by the parameters a model stores. Following standard Transformer compute accounting~\cite{kaplan2020scaling,hoffmann2022training},
\begin{equation}
    \mathrm{FLOPs/token}\approx 2N_{\mathrm{active}},
    \label{eq:flops-per-token}
\end{equation}
where $N_{\mathrm{active}}$ denotes the parameters engaged in the forward pass; for sparse mixture-of-experts models only a subset of experts is routed per token, so per-token computation stays approximately constant as the expert pool grows~\cite{fedus2022switch}. Activated parameters therefore index per-token computation up to a constant factor, though not serving cost in full, which also reflects attention implementation, key--value cache size and deployment stack. Panels \textbf{E}--\textbf{H} of \Cref{fig:parameter-scaling-row} adopt this view, fitting dense and mixture-of-experts checkpoints separately (dashed and dotted): dense checkpoints keep their position from \textbf{A}--\textbf{D}, whereas the MoE checkpoints move left according to their activated expert count and
retain much of their accuracy advantage at lower effective compute,
suggesting that, where education systems must serve many concurrent learners under per-token cost constraints, sparse architectures can offer favourable accuracy--compute tradeoffs in the evaluated families:
the 35B sparse thinking configuration activates only 3B parameters yet exceeds
the 9B dense checkpoint in foundational literacy and knowledge
(93.8\% versus 91.1\%) and subject-specific pedagogical expertise
(90.4\% versus 86.0\%), while essentially matching it on pedagogical
principles (89.2\% for both).

 \subsection{Effect of Inference Scaling}
The preceding analyses treated the thinking and no-thinking configurations of a
model as separate entries, which reports whether extended reasoning helps on
average but not where it helps.
We therefore compare the two configurations group by group, relating the
generated length of each to the score it achieves.
·\paragraph{Definition of inference scaling.}
Let \(m\in\{\mathrm{T},\mathrm{N}\}\) index the thinking and no-thinking
configurations of the same model, and let \(i\) index an evaluation item.  We
measure the generated length in characters as
\begin{equation}
    \ell_{mi}
    =
    \left|\operatorname{reasoning}_{mi}\right|
    +
    \left|\operatorname{response}_{mi}\right|,
    \label{eq:inference-generated-length}
\end{equation}
where the two terms are the hidden reasoning trace and the emitted model
response, respectively.  The corresponding item result \(r_{mi}\in[0,100]\) is as
defined in \Cref{sec:evaluation}.

To make the analysis independent of any particular taxonomy, we use the grouping
notation of \Cref{eq:group-score}: for each model
configuration and group, the quantities plotted in the grouped comparison are
\begin{equation}
    \bar{\ell}_{mg}
    =
    \frac{1}{|\mathcal{I}_{mg}|}
    \sum_{i\in\mathcal{I}_{mg}}\ell_{mi},
    \qquad
    \bar{r}_{mg}
    =
    \frac{1}{|\mathcal{I}_{mg}|}
    \sum_{i\in\mathcal{I}_{mg}}r_{mi}.
    \label{eq:inference-group-means}
\end{equation}
Thus, Panel \textbf{A} shows the item-level pairs
\((\ell_{mi},r_{mi})\), whereas Panel \textbf{B} shows
\((\bar{\ell}_{mg},\bar{r}_{mg})\); its dashed segment joins the two model
configurations evaluated under the same group \(g\).

\begin{figure}[H]
\centering
\includegraphics[
  width=\linewidth,
  height=0.72\textheight,
  keepaspectratio
]{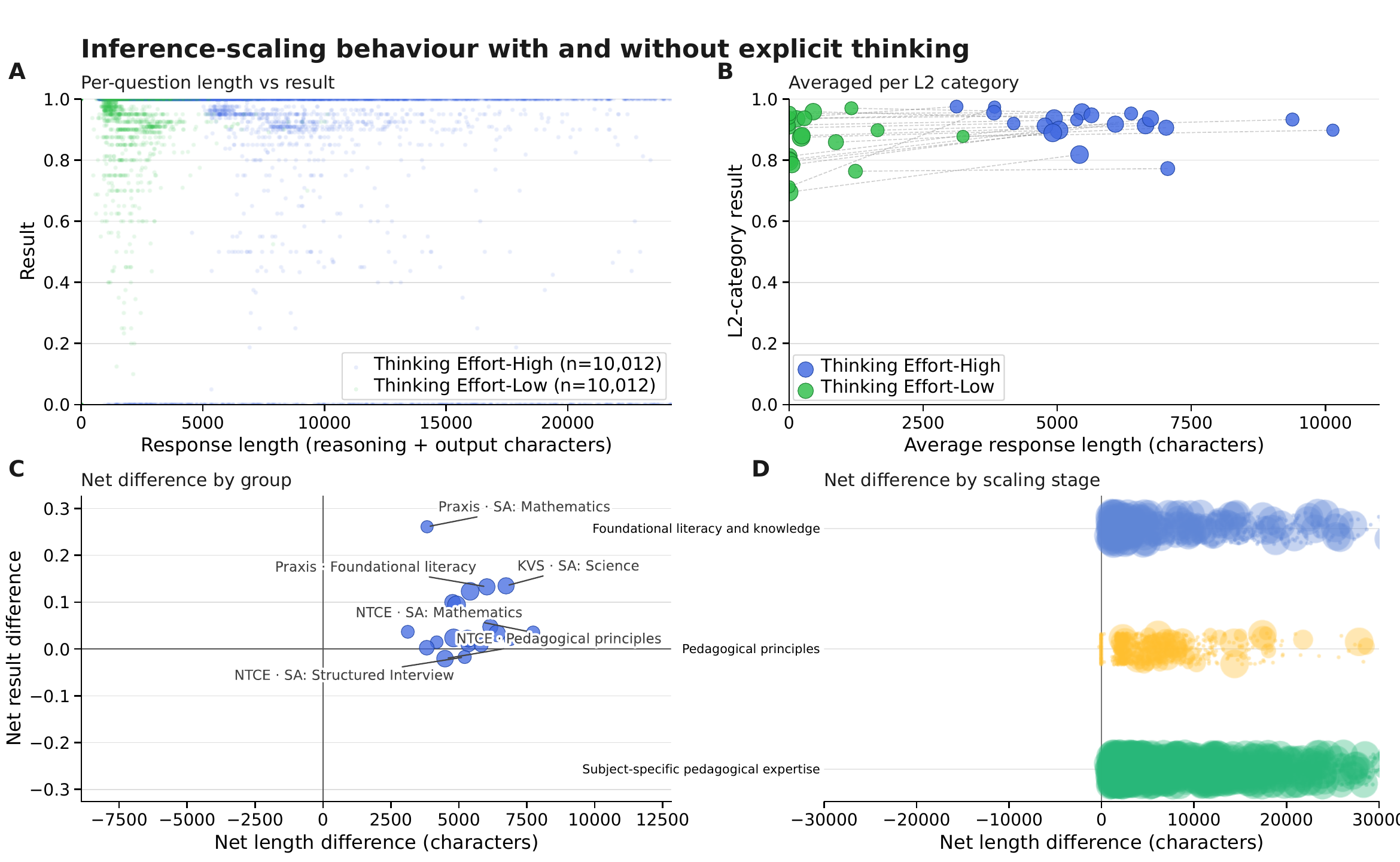}
\caption{Inference-scaling behavior for Qwen3.6-35B with and without explicit
thinking. \textbf{A}, per-question response length versus result. \textbf{B},
the same data averaged per aligned L2 category, with dashed connectors joining
the two variants of each category. \textbf{C}--\textbf{D}, paired
thinking-minus-no-thinking differences in length and result, averaged per
group (\textbf{C}) and stratified by the S1/Core, S2/PLT, and
S3/Subject/India stages (\textbf{D}). In \textbf{D}, marker area (points
squared) is
\(\displaystyle
A_i=10+520
\left[
\max(0,\Delta r_i)/\max(G,10^{-9})
\right]^{0.75}
\),
where \(G=\max_j\max(0,\Delta r_j)\).}
\label{fig:inference-scaling-output-length}
\end{figure}

Net inference-scaling effects are computed only from matched items.
Net inference-scaling effects are computed only from matched items. Let
\(\mathrm{T}\) and \(\mathrm{N}\) denote the thinking and no-thinking
configurations of the same model, respectively.
For every \(i\in. \mathcal{I}_{\mathrm{T}g} \),
\begin{equation}
    \Delta\ell_i=\ell_{\mathrm{T}i}-\ell_{\mathrm{N}i},
    \qquad
    \Delta r_i=r_{\mathrm{T}i}-r_{\mathrm{N}i}.
    \label{eq:inference-item-differences}
\end{equation}
The group-level net differences are the unweighted means of these paired
item-level differences,
\begin{equation}
    \overline{\Delta\ell}_{g}
    =
    \frac{1}{|\mathcal{I}_{\mathrm{T}g}|}
    \sum_{i\in\mathcal{I}_{\mathrm{T}g}}\Delta\ell_i,
    \qquad
    \overline{\Delta r}_{g}
    =
    \frac{1}{|\mathcal{I}_{\mathrm{T}g}|}
    \sum_{i\in\mathcal{I}_{\mathrm{T}g}}\Delta r_i.
    \label{eq:inference-group-differences}
\end{equation}
Panel \textbf{C} plots
\((\overline{\Delta\ell}_{g},\overline{\Delta r}_{g})\) for each chosen group.
Panel \textbf{D} retains the individual paired observations
\(\Delta\ell_i\), stratifies them by the broader certification-stage grouping,
and uses marker area to encode the positive score gain
\(\max(0,\Delta r_i)\); it does not replace the item-level differences with a
stage mean.  In the current figure, Panels \textbf{B} and \textbf{C} use the
aligned system--L2 categories (with the aligned Core and PLT categories retained
as whole stages), while Panel \textbf{D} uses the three aligned certification
stages.
\Cref{fig:inference-scaling-output-length} shows this analysis for Qwen3.6-35B,
one of the models evaluated in both configurations and situated at the locally
deployable scale where the accuracy--cost tradeoff is most consequential.
Extended reasoning helps in 18 of the 20 groups, and
the two groups where it does not are both assessments of pedagogical judgement
rather than of subject content, a pattern that sets them apart from every other
category.
Both lose ground despite generating substantially more text: NTCE pedagogical
principles falls 2.1 points across 4{,}491 additional characters and the
structured interview 1.7 points across 5{,}216, while Praxis pedagogical
principles gains only 0.9 points for 5{,}823.
IN contrast, Praxis mathematics gains 26.1 points from 3{,}831 additional
characters---the expected behaviour on tasks with verifiable intermediate steps.

\subsection{Discipline Breakdown}
A tier-level gap of a few points could arise from a uniform deficit spread over
every competence or from large deficits confined to a few. To distinguish these,
we compare the leading system of each access tier on all 20 L1-by-L2 subsets.
Using the grouping notation of \Cref{eq:group-score}, we let \(g=c\) denote
one of the 20 displayed L1-by-L2 subsets and let \(\gamma(i)=c\) assign each
item to its corresponding subset. Thus,
\(\mathcal{I}_{mc}=\{i:\gamma(i)=c\}\) contains the items in subset \(c\)
for which model \(m\) has a valid result, and the displayed score is
\(A_{mc}\) as defined in \Cref{eq:group-score}.
Panel \textbf{a} shows that this imbalance is not a simple function of model
class. The single-accelerator Qwen3.6 27B, a newer training generation positioned
primarily for agentic use~\cite{qwen2026qwen36}, matches or exceeds both larger
models wherever the task is reasoning-dominant: it scores
100.0\% on Praxis mathematics against 95.7\% and
96.5\% for the proprietary and industrial-scale open-weight models,
respectively; 95.1\% on Chinese mathematics against
94.7\% and 93.1\%; and leads on Chinese foundational
literacy (97.0\%) and Praxis language
arts (94.9\%). Its deficits are concentrated instead
where breadth of memorized, region-specific knowledge is required---
82.0\% on Indian language arts against the 98.8\%
frontier, and 7.1--8.2 percentage points behind the
frontier on Indian social science and science. Single-accelerator models therefore
appear to close the reasoning gap well before the knowledge-coverage gap.

\textbf{Merging matched categories across countries.}
For a harmonized cross-country category \(h\), let
\(\mathcal{G}_h\) denote the set of country-specific groups mapped to \(h\),
with at most one group from each represented country. Using the group scores
defined in \Cref{eq:group-score}, the cross-country panel reports
\[
    A^{\mathrm{macro}}_{m,h}
    =
    \frac{1}{|\mathcal{G}_h|}
    \sum_{g\in\mathcal{G}_h} A_{m,g}.
\]
Thus, each represented country contributes one group score and receives equal
weight irrespective of its number of evaluation items.
The merged view preserves this ordering: the single-accelerator model stays within
2.6 percentage points of the larger-model frontier on teaching
demonstration, pedagogical principles and foundational literacy; on mathematics it
instead leads, with 97.5\% versus 95.2\%. It falls
3.9--5.9 points behind the frontier on
social science, science and language arts. The single-accelerator model therefore
approaches the frontier more closely on core professional competences than on breadth
of disciplinary and language-specific knowledge.

\begin{figure*}[!t]
\centering
\includegraphics[
  width=\textwidth,
  keepaspectratio
]{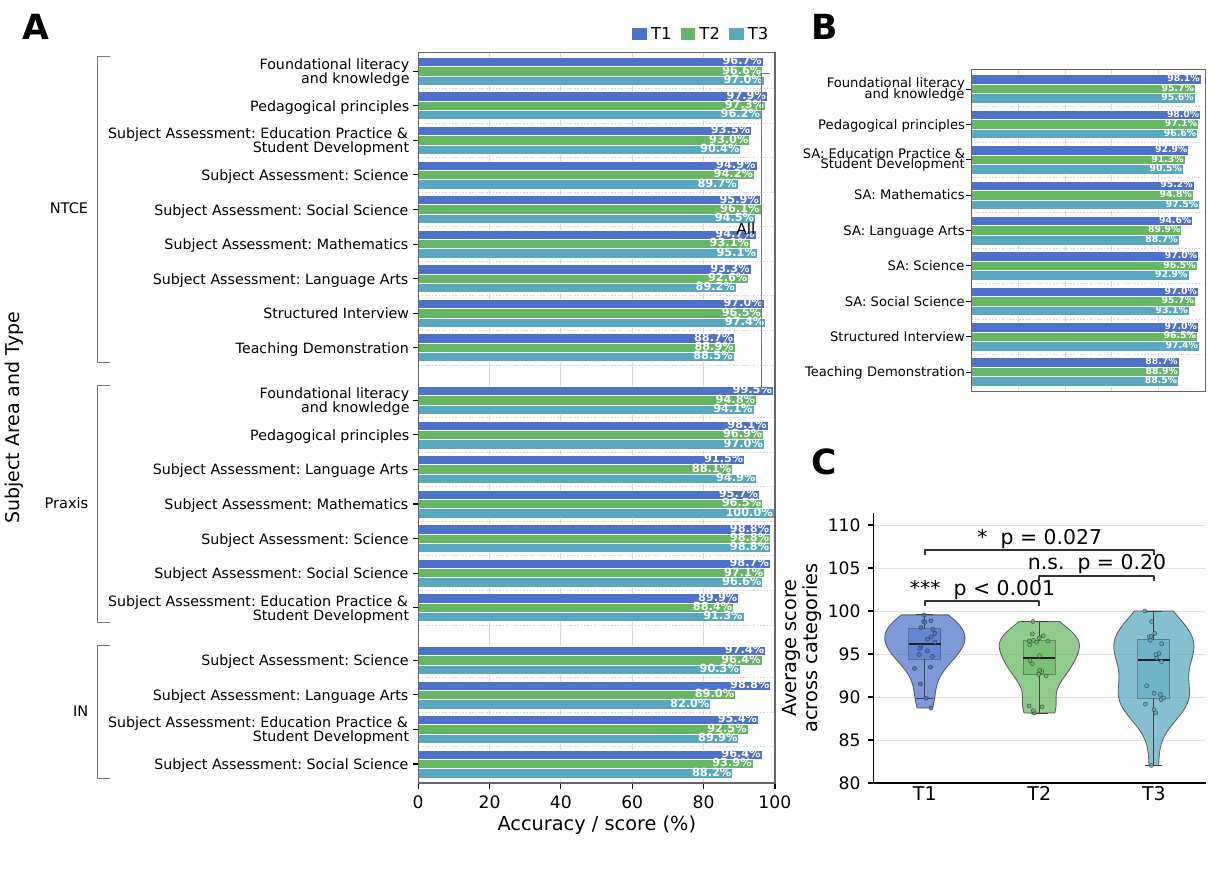}
\caption{\footnotesize\textbf{Performance across examination subjects.}
  Distribution of 20 paired category scores for the best closed-source, open-source and single-accelerator models. Pairwise $p$ values use two-sided Wilcoxon signed-rank tests: for $d_i=A_{r,i}-A_{s,i}$, exact zeros are discarded, and the average ranks $\rho_i$ of $|d_i|$ form $R_{\pm}=\sum_{d_i\gtrless0}\rho_i$ and $W=\min(R_+,R_-)$.
}
\label{fig:mmlu-style-breakdown}
\end{figure*}
\textbf{Weaknesses shift with model class.}
Because the 20 subsets are matched across models, every pair of model classes
yields 20 paired differences $d_i$. Scores are bounded and their differences are
not normally distributed, so we compare classes with two-sided Wilcoxon
signed-rank tests on these differences rather than with $t$-tests; the statistic
is defined in the caption of \Cref{fig:mmlu-style-breakdown}.
The location of each model's knowledge gaps is systematic rather than random.
These patterns are consistent with a knowledge-coverage rather than a reasoning bottleneck for single-accelerator models.

\subsection{Error Reason Analysis}
\label{app:error-reason-findings}
\begin{figure}[!t]
\centering
\includegraphics[
  width=\linewidth,
  height=0.55\textheight,
  keepaspectratio
]{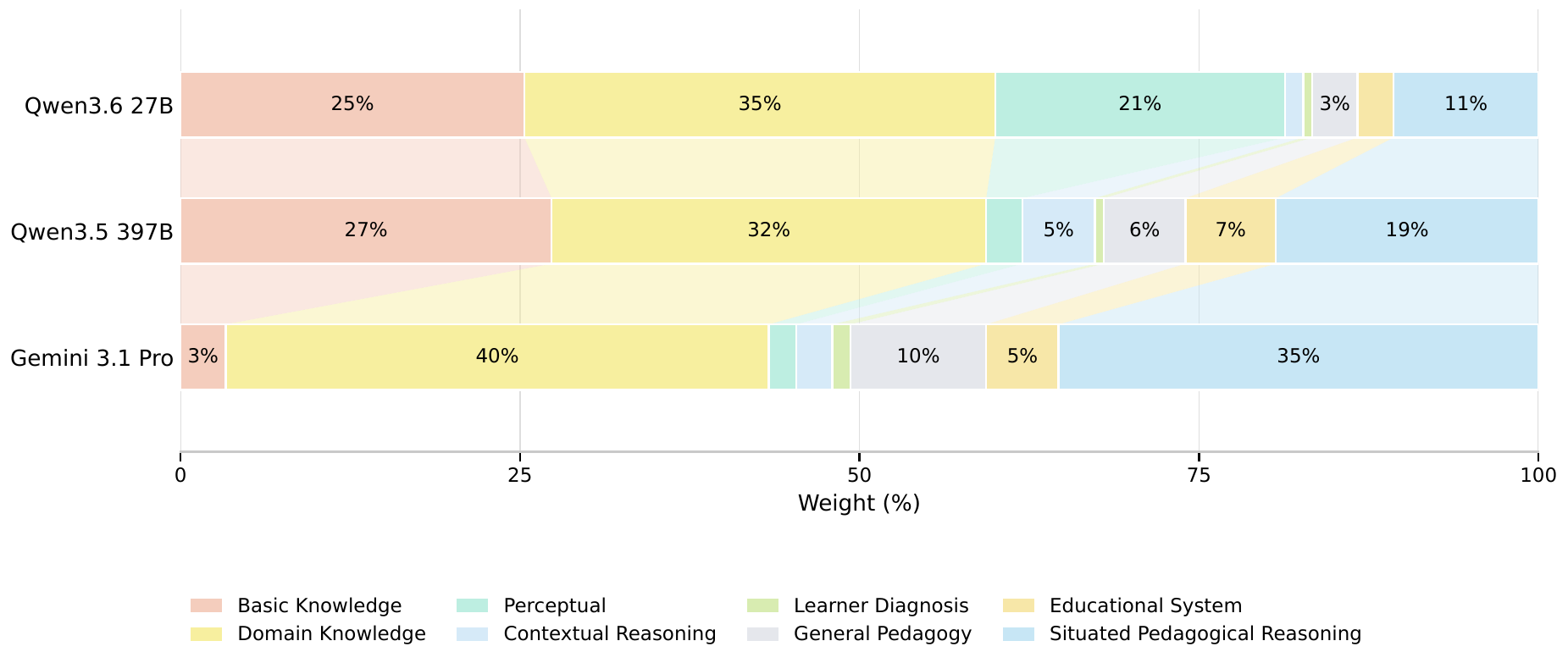}
\caption{Distribution of annotated error samples across the error-reason taxonomy.}
\label{fig:error-sample-distribution}
\end{figure}

Scores locate the remaining gaps but do not explain them: a model that loses
points on subject-specific pedagogy may be missing the disciplinary content, the
instructional principle, or the learner it is asked to reason about.
To separate
these causes we follow the error-analysis methodology of
MMMU~\cite{yue2024mmmu},  and annotate the primary failures mode
of the leading model in each access tier.
For each tier leader---Gemini 3.1 Pro (T1), Qwen3.5-397B (T2) and Qwen3.6-27B
(T3)---we meticulously examine 150 randomly sampled error instances drawn from
its incorrect selected responses across the three examinations.
Expert annotators identify the root cause of each misprediction against the
reference answers and explanations, using an eight-category taxonomy:
four non-pedagogical categories---Basic Knowledge (E1), Domain Knowledge (E2), Perceptual (E3), and Contextual Reasoning (E4) errors---and
four pedagogy-related categories---Learner Diagnosis (E5), General Pedagogical (E6), Educational System (E7), and Subject-specific Pedagogical (E8) errors.
For each model we examine 150 incorrect responses,
The resulting distribution is shown in \Cref{fig:error-sample-distribution}.

We observe that scaling shifts the error frontier from knowledge acquisition toward situated pedagogical reasoning
Capability removes the recall and perception failures---together 46.6\% of the
single-accelerator model's errors against under 6\% of the frontier model's---
but not the two residuals that certification actually tests: the four
pedagogy-related categories (E5--E8), which rise from 17.3\% of the single-accelerator
model's errors to 52.0\% of the frontier model's, and Domain Knowledge, still the
largest single category at every tier (32.0--40.0\%).
What survives scaling therefore lies at the core of professional educational competence: mastery of the subject taught, and its translation into instruction---the primary direction for future work.

\subsection{Performance on the EDU-Verse Benchmark}
High average accuracy is not the quantity that governs educational deployment: a
model that answers 96.8\% of selected-response questions correctly still fails on
one item in thirty, and those failures are what a classroom would encounter.
To keep the
instrument informative as models improve, we construct a harder subset from
precisely those failures:
from the questions that at least one of the T1, T2 and leading T3 systems answers
incorrectly, experts take  uniformly sampled residual questions per system and all
of the overall leader's errors, providing a gold verdict and explanation for each item,
yielding a curated residual-difficulty set of 1,050 unique questions spanning all three countries (502 KVS, 370 NTCE, and 178 Praxis).

The two views differ in resolution rather than in what they measure. On the full
set, systems that differ substantially in educational competence are compressed
into a few points of accuracy; on the residual set the same systems separate by
tens of points, and their ordering changes. Residual difficulty is therefore the
component of the benchmark that still carries information about model quality.

\begin{figure}[!t]
\centering
\includegraphics[width=\linewidth]{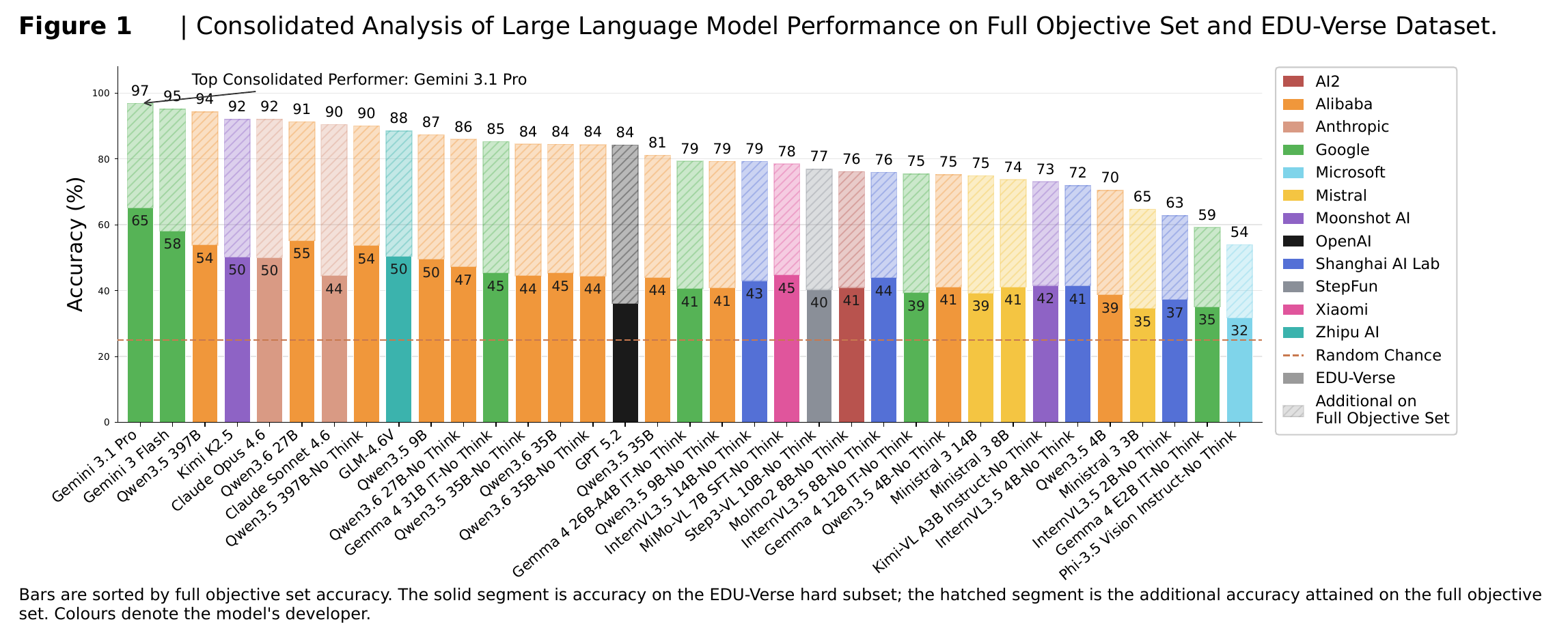}
\caption{Every evaluated model's accuracy on EDU-Verse (solid segment) and the additional accuracy it attains on the full selected-response set (hatched segment), sorted by full-set accuracy. Colours denote the model's developer; the dashed line marks random-choice accuracy.}
\label{fig:edupro-unioned-selected-response-model-scores}
\end{figure}

\section{Discussion}\label{sec:discussion}

\paragraph{Accessibility expands as open models approach frontier performance}
The strongest proprietary model, Gemini 3.1 Pro, reaches a weighted overall score
of 96.4\%, against 94.1\% for the strongest open-weight system, Qwen3.5-397B, and
91.1\% for Qwen3.6-27B, which runs on a single accelerator. Each step down the
access ladder therefore costs a few points---roughly 220 and 500 of the 10,012
evaluated items---against a roster that spans 44 points end to end, and the
differences between tiers are smaller than the variation within them. Parameter
count is likewise not a sufficient explanation of educational capability:
Qwen3.5-9B and Qwen3.5-35B reach 86.9\% and 90.6\%, and the activated-parameter
analysis shows that mixture-of-experts systems retain much of their accuracy while
computing with a fraction of their stored weights.
For reference, these examinations are selective for the humans who sit them: the
national NTCE is reported to pass approximately 25\% of
candidates~\cite{you2022professionalisation}. Results are issued on scales
unlike ours---a 100--200 scaled score for Praxis, where medians run 162--175
across the Core and PLT assessments~\cite{ets_praxis_scores_2022}, and a
120-point converted score for the NTCE, anchored so that 70 marks the passing
line for every subject and session~\cite{neea_ntce_score_2026}. These
distributions provide context for the magnitudes reported here, but do not
support converting a model's score rate into a certification outcome.

\paragraph{Residual evaluation reveals hidden capability gaps}
Aggregate accuracy increasingly measures what models have already mastered rather
than what separates them.
Two categories have stopped discriminating altogether: the three leading systems
sit within 0.9 points of one another on the structured interview (96.5--97.4\%)
and within 0.4 points on the teaching demonstration (88.5--88.9\%),
though the annotating experts reported that the responses themselves remained
substantively varied, with no sign of collapose onto common patterns---a uniformity that may reflect both the
fluency of current models and the tendency of rubric-based judges to reward
well-formed answers.
Restricting the comparison to residual difficulty restores
the resolution the aggregate loses: on EDU-Verse the spread among leading models
widens from 12.6 to 37.1 points, and their ordering changes.

\paragraph{Pedagogical reasoning remains the frontier beyond scaling}
The same gap appears whichever way we vary the models. Scaling, activated compute,
extended reasoning and a newer training generation each improve foundational
competences faster than subject-specific teaching ability, which remains the
lowest-scoring category in every view and the largest source of residual error in
the strongest models.
All three leading systems clear
95\% on foundational literacy and on general pedagogical principles, yet each
gives way somewhere among the five subject assessments: the proprietary system
falls to 92.9\% on education practice and student development, and the
leading open-weight and single-accelerator systems to 89.9\% and 88.7\% on language arts.
Pedagogical content knowledge---making particular subject
matter teachable to particular learners---is therefore not a by-product of scale,
compute or deliberation, but a capability that will have to be trained and
measured directly.

\section{Limitations}\label{sec:limitation}

\textbf{Current scope.}
This paper reports results on the NTCE written modules (S1--S3), structured interviews, teaching demonstrations, the U.S.\ Praxis examination series, and India's KVS teacher-recruitment examinations (MCQ only).
Evaluation of adversarial moral dilemma extensions and multi-agent classroom simulation is underway and will be reported in a subsequent version.
Results for the French CAPES are in preparation.

\textbf{Broader implications.}
EDU 1.0 does not advocate replacing human teachers with LLMs. Rather, by applying the same rigorous standards used to certify human teachers, we aim to provide a principled framework for understanding where LLMs can effectively support education and where critical gaps remain.

\bibliography{main}
\bibliographystyle{abbrvnat}

\end{document}